\documentclass[final]{ws-ijmpa}
\usepackage[super,compress]{cite}
\usepackage{amsmath}
\usepackage{mathrsfs}
\usepackage{amssymb,amsfonts}
\usepackage{bm,bbm}
\usepackage{graphicx}
\usepackage{esint}
\usepackage{color}
\usepackage[breaklinks]{hyperref}
\usepackage[all]{xy}
\usepackage{arydshln} 

\let\normalcolor\relax  

\newcommand{\be}{\begin{equation}}
\newcommand{\ee}{\end{equation}}
\newcommand{\ba}{\begin{eqnarray}}
\newcommand{\ea}{\end{eqnarray}}

\def\bs{}
\def\es{}

\def\a{\alpha}
\def\b{\beta}
\def\de{\delta}
\def\g{\gamma}
\def\la{\lambda}

\def\G{\Gamma}

\def\vr{\varrho}

\def\N{\nabla}

\def\cD{\mathcal{D}}

\def\cF{\mathcal{F}}

\def\cH{\mathcal{H}}

\def\cK{\mathcal{K}}

\def\cM{\mathcal{M}}

\def\cO{\mathcal{O}}

\def\cR{\mathcal{R}}

\def\cV{\mathcal{V}}
\def\bE{\mathbbm{e}}

\def\ds{d_{\rm S}}
\def\dh{d_{\rm H}}

\def\p{\partial}

\def\B{\Box}
\newcommand{\Eq}[1]{(\ref{#1})}
\def\com{\color{magenta}}
\def\cob{\color{blue}}

\newcommand{\oarX}[1]{\href{http://arxiv.org/abs/#1}{arXiv:{\com#1}}}
\newcommand{\arX}[1]{\href{http://arxiv.org/abs/#1}{arXiv:{\com#1}}}
\newcommand{\doin}[6]{\href{http://dx.doi.org/#1}{\cob {\it #2 #3} {\bf #4}, #5 (#6)}}
\newcommand{\doij}[5]{\href{http://dx.doi.org/#1}{\cob {\it #2}    {\bf #3}, #4 (#5)}}
\newcommand{\ndoin}[6]{\href{#1}{\cob {\it #2 #3} {\bf #4}, #5 (#6)}}
\newcommand{\tia}[1]{}

\def\rme{e}
\def\rmd{d}
\def\rmi{i}

\newcommand{\m}{\mu}
\newcommand{\n}{\nu}

\begin{document}


\title{MULTIFRACTIONAL SPACETIMES, ASYMPTOTIC SAFETY AND HO\v{R}AVA--LIFSHITZ GRAVITY}
\author{GIANLUCA CALCAGNI}
\address{Instituto de Estructura de la Materia, IEM-CSIC, Serrano 121, 28006 Madrid, Spain\\ calcagni@iem.cfmac.csic.es}

\maketitle

\begin{abstract}
We compare the recently formulated multifractional spacetimes with field theories of quantum gravity based on the renormalization group (RG), such as asymptotic safety and Ho\v{r}ava--Lifshitz gravity. The change of spacetime dimensionality with the probed scale is realized in both cases by an adaptation of the measurement tools (``rods'') to the scale, but in different ways. In the multifractional case, by an adaptation of the position-space measure, which can be encoded into an explicit scale dependence of effective coordinates. In the case of RG-based theories, by an adaptation of the momenta. The two pictures are mapped into each other, thus presenting the fractal structure of spacetime in RG-based theories under an alternative perspective.
\end{abstract}

\date{September 19, 2012}



\ccode{PACS numbers: 04.60.-m, 05.45.Df, 11.10.Gh, 11.10.Kk}
\

\centerline{\doin{10.1142/S0217751X13500929}{Int.\ J.\ Mod.\ Phys.}{A}{28}{1350092}{2013} [\arX{1209.4376}]}


\section{Introduction}\label{intro}

When geometry becomes quantum, the dimension of effective spacetime changes with the scale. This phenomenon has been dubbed dimensional reduction \cite{tHo93,Car09} 
 or dimensional flow \cite{fra1}, and it appears across a variety of approaches both in the continuum (such as in noncommutative spacetimes, asymptotic safety, Ho\v{r}ava--Lifshitz gravity and fractal field theory) and in discrete settings (causal dynamical triangulations, spin foams and loop quantum gravity among others). While in the latter category of models it is often problematic to get the continuum limit of a discrete geometry, the former class can be manipulated more easily and it is particularly suitable to make contact with physics and extract some phenomenology across different scales.

Multiscale phenomena are not a foreign topic in quantum field theory thanks to the massive deployment of renormalization group (RG) techniques in the most various contexts. When geometry itself is quantized, however, dimensional reduction acquires certain universal traits and both its underlying mechanism and the precise relation with the RG flow need to be explored thoroughly. This problem can be naturally studied in fractal field theory \cite{fra1,fra2,fra3}, especially in its fractional incarnation \cite{fra4,frc1,frc2,frc3,fra6,frc4,frc5} (see Ref.\ \citen{AIP} for a review). In this case, one can easily tune the ingredients of a Lebesgue--Stieltjes fractal field theory of fixed dimensionality living in fractal Minkowski spacetime $\cM_v^D$, the latter being embedded in $D$-dimensional ordinary Minkowski spacetime $M^D$. Namely, these ingredients are: (i) the measure weight $v(x)$ in position space and, from that, the Hausdorff dimension $\dh(\cM_v^D)$ of position space \cite{frc1,frc2}; (ii) the measure weight $w(p)$ in momentum space and, from that, the Hausdorff dimension $\dh(\cM_w^D)$ of momentum space \cite{frc3,frc4}; (iii) the Laplacian operator $\cK$ in position space \cite{frc2,frc3,frc4}; (iv) the type of diffusion (normal or anomalous) associated with $\cM_v^D$ \cite{frc4}. From (i), (iii) and (iv), one can extract (v) the spectral dimension $\ds(\cM_v^D)$ of spacetime \cite{frc1,frc4}. This geometry can be extended to a multiscale setting \cite{frc2,frc4} and fields can be introduced thereon. Dimensional flow is fully analyzable and one can exercise quantitative control at all scales, not just in isolated regimes along the flow \cite{frc2,fra6}. Also, both at different points in the flow and as a whole, diffusion in quantum spacetime is given a physical and probabilistic interpretation in terms of stochastic processes \cite{frc4}.

Although fractal spacetimes and field theories thereon have been proposed as fundamental, an alternative point of view is to regard them as \emph{effective models} describing certain (regimes of) other, independent quantum-spacetime/quantum-gravity theories. The strategy is to identify the above five data in the fundamental theory we are interested in. In the known cases, we do not have all of them at hand. In theories defined on group manifolds (e.g., noncommutative spaces), an additional complication is to extract analogs of (i)--(v).\footnote{A wide class of noncommutative spacetimes can be mapped to fractional geometries by recognizing that the classical cyclicity-invariant measure in position space, associated with a given (possibly nonlinear) spacetime algebra, is of fractional type \cite{ACOS}. In particular, this helped in reinterpreting the problematic cyclic measure $\rmd {\bf x}/(\prod_i x_i)$ of $\kappa$-Minkowski spacetime as the asymptotic limit of a more complicated geometry.} For theories living in a continuum, such as asymptotic safety (also known as quantum Einstein gravity, QEG \cite{Wei79,Reu1}; see Refs.\ \citen{Nie06,Lit11,ReS11,RSnax} for reviews), this difficulty is partially removed but some of the data may be missing or undecided. (In the case of QEG, (i) is known at any scale, (ii) is assumed, (iii) is tailored compatibly with the RG flow, (v) is known at the UV and IR fixed points, and (iv) can be inferred consequently \cite{CES}.) Next, we match these data with the corresponding ones of the fractional theory. Even when both models are on the continuum and on $c$-numbers, this step is not always just a matter of recasting the description of the fundamental theory in terms of the language of fractal geometry. Rather, fractal geometry can give constructive insight to fill the gaps in the fundamental theory, remove ambiguities in previous guesswork, and clarify the physical meaning of the ingredients. Last, one extends the mapping to multifractal geometries where the effective spacetime dimension varies with the scale.

Due to the similarity of dimensional flow in QEG, multifractional spacetimes and Ho\v{r}ava--Lifshitz (HL) field-theory approach \cite{Hor2,Hor3}, one may ask whether the multiscale (in particular, multifractal) geometry of multifractional spacetimes can serve as an alternative or complementary description of dimensional flow in these and other RG-based theories. The answer is in the affirmative. In the present paper, we gain some insight on the nature of the RG flow when certain quantities (the metric in QEG, position coordinates in HL gravity) exhibit anomalous scaling,
 by interpreting a multiscale geometry as a continuous hierarchical adaptation of ``rods'' (i.e., geometry measurements) on the length scale of the probed phenomena, with respect to the scale of classical macroscopic rods. Thus, dimensional reduction stemming from RG techniques admits a ``dual'' description in terms of multifractal geometry.

It is important to stress that the scope of this dual interpretation is limited by the physical assumptions, i.e., by the symmetries of the system. In particular, from the symmetry choice in the action different Laplacians lead to physically inequivalent models. In other words, we concentrate on some generic features of position and momentum space, ignoring dynamics. Even if this mapping eventually amounts to a different \emph{parametrization} of the theories, it does highlight a \emph{physical} property (anomalous scaling) of the latter, because it involves momentum, a parameter with direct physical significance in experiments.

We take the three examples of multifractional spacetimes, QEG and HL gravity for illustration, in the absence of curvature (the metric is the Minkowski one, $g_{\mu\nu}=\eta_{\mu\nu}={\rm diag}(-,+,\cdots,+)$). In all three cases, the Laplace--Beltrami operator $\cK$ converges in the infrared (IR) to the usual d'Alembertian,
\be\label{lbo}
\cK\ \stackrel{\rm IR}{\sim}\ \B=-\p_t^2+\N^2\,,
\ee
where $\N^2$ is the spatial Laplacian. At scales between the IR and the ultraviolet (UV), we have the following limits.
\begin{itemize}
\item {\it Multifractional spacetimes}. To get a two-dimensional geometry in the UV \cite{frc2}, one can choose the second-order self-adjoint operator
\be\label{ka}
\cK_\a=\eta^{\mu\nu}\cD_\mu\cD_\nu\,,\qquad \cD_\mu:=\frac{1}{\sqrt{v_\a(x)}}\,\p_\mu\left[\sqrt{v_\a(x)}\,\,\cdot\,\right]\,,
\ee
where $v_\a(x)$ is the spacetime measure weight at a given scale governed by the parameter $\a$. In the UV, $\a=2/D$, where $D$ is the number of topological dimensions. This is a particular model of multifractional spacetimes. Poincar\'e invariance is broken except in free models \cite{frc6}. Renormalization can be studied in a perturbative quantum field theory context \cite{frc2}.
\item {\it Asyymptotic safety}. Lorentz invariance is preserved throughout the RG/dimensional flow and, at any given scale, in early versions the Laplacian was effectively given by
\be\label{effas0}
\cK_\de \sim \B^{1+\frac{\de}{2}}\,,
\ee
where $\de=2$ at the UV non-Gau\ss{}ian fixed point, $\de=0$ in the IR and it can acquire other nontrivial values in between. Equation \Eq{effas} is meant to be only qualitative, since the actual scale identification takes place at the level of the metric, not of the Laplacian, and Eq.\ \Eq{effas} is more likely to become the usual Laplace--Beltrami operator, while the diffusion-equation operator is modified nontrivially \cite{CES}:
\be\label{effas}
\cK_{\normalfont\textsc{qeg}} = \B\,.
\ee
The theory is studied nonperturbatively via the functional RG approach.
\item {\it Ho\v{r}ava--Lifshitz spacetimes}. Here diffeomorphism invariance is preserved only on spatial slices and along the time direction, which is singled out as preferred. The Laplace--Beltrami operator is then
\be
\cK_z = -\p^2_t+\N^{2z}\,,
\ee
where $z$ is a parameter fixed at $z=D-1$ in the UV and, effectively, $z=1$ in the IR, Eq.\ \Eq{lbo}. As in multifractional models, renormalization is formulated perturbatively.
\end{itemize}
Even if these models are characterized by a similar or identical dimensional flow, they are essentially different in other predictions. However, the Laplacian operator does not enter the geometric description presented in this paper, which is rather general. Therefore, the latter may be regarded not as a physical duality between theories, but as a complementary geometric description of their multiscale properties related to momentum space.

In Sec.\ \ref{sec2}, we briefly review multifractional, QEG and Ho\v{r}ava--Lifshitz spacetimes. Section \ref{mfm} is the core of the paper. We shall concentrate on the comparison between multifractional theories and asymptotic safety; the case of Ho\v{r}ava--Lifshitz stems from a slight generalization of the obtained results. A discussion follows in Sec.\ \ref{disc}.


\section{Three Approaches to Quantum Geometry}\label{sec2}


\subsection{Multifractional spacetimes}\label{revmf}

Without loss of generality we concentrate on fractional Minkowski spacetime $\cM_v^D=\cM_\a^D$, where the metric is $\eta_{\mu\nu}$ and there is no gravity. $\cM_\a^D$ lives in $D$-dimensional ordinary Minkowski spacetime $M^D$, (spanned by coordinates $x^\mu$, $\mu=0,1,\dots,D-1$) but, contrary to the latter, it is endowed with a nontrivial measure of the form
\bs\ba
\rmd\vr_\a(x) &=&\rmd^Dx\,v_\a(x)\,,\\
v_\a(x) &=&\prod_\mu v_\a(x^\mu)\label{fac}\\
&:=&\prod_\mu \frac{|x^\mu|^{\a_\mu-1}}{\Gamma(\a_\mu)}\,,
\ea\es
where $\G$ is the gamma function and $\a_\mu$ are $D$ real parameters (``fractional charges'') in the (further restrictable) range $0<\a_\mu\leq 1$. From the scaling law of the measure, $\vr_\a(\la x)=\la^{\dh}\vr_\a(x)$, one infers that the Hausdorff dimension of fractional spacetime is $\dh=\sum_\mu\a_\mu$, as one can show also with other methods \cite{frc1}. One is to calculate (in Euclidean signature) the volume of a $D$-ball of radius $R$,
\be\label{volD}
\cV^{(D)}(R) = \int_{D{\rm-ball}} \rmd\vr_\a(x) \propto R^{\sum_\mu\a_\mu}\,,
\ee
where the exponent is by definition the Hausdorff dimension of space. The simplest type of fractional measure is ``isotropic'' and $\a_\mu=\a$ are all equal, so that $\dh=D\a$; we shall focus our attention on this configuration except when discussing Ho\v{r}ava--Lifshitz spacetimes. It is also possible to extend the geometric construction to complex-valued fractional charges \cite{fra4,frc2,ACOS} or more general factorizable measures \cite{frc5,frc6} but we shall not do so since scaling properties are more apparent in the real-order fractional case.

The measure can be reduced to the Lebesgue one by defining a set of coordinates, which we call ``geometric,'' such that
\be\label{geo}
q^\mu:=\vr_\a(x^\mu)=\frac{{\rm sgn}(x^\mu)|x^\mu|^\a}{\Gamma(\a+1)}\,,
\ee
so that $\rmd\vr_\a=\rmd^Dq$. Geometric coordinates will be essential to compare fractional theories with other proposals. Obviously a change of variables does not modify the geometry. Consider one direction, $\mu$ fixed, and the quantity $\cR = q(R) = \int_0^{q(R)}\rmd q= \int_0^R \rmd x\, v_\a(x) \propto R^\alpha$. The volume element \Eq{volD} can be also expressed as $\cV^{(D)} \propto R^{\alpha D} = \cR^D$, but from this one should not surmise that $\dh=D$. In fact, $\cR$ does not correspond to the fractional generalization of distance, since the variables $q$ have anomalous scaling $q(\la x)$ = $\la^\alpha q(x)$ (their momentum units are $[q]=-\a$ because $[x]=-1$ by definition). Hence, $\text{(volume)} = \text{(distance)}^{D\a}$. It is easy to meet with contradictions if one confuses the measure structure with a traditional metric structure, and forgets about the anomalous scaling property of the coordinates $q$, which is a consequence of a nonstandard definition of momentum space.

The paper revolves around this last point. In fractional spaces, even in the absence of curvature effects, geometry is not measured by standard rods (i.e.\ an ordinary line element with metric). A fractional length at a scale where the $\alpha$th component of the multiscale measure dominates is measured in ordinary integer length units to the power of $\alpha$. In fractal geometry one encounters a similar situation. Loosely recalling the example given by Mandelbrot in his seminal 1967 paper \cite{Man67}, if one uses traditional length rods to measure the coast of Great Britain, one finds that the result is infinity. If one uses area rods, one finds zero, because it does not fill the whole plane. To get a finite result for a fractal curve, one should not measure its geometry by ordinary (integer, length or area) units. It is well known that the Hausdorff measure $\cH_s$ with parameter $s$ can be also 0 or $\infty$ at the critical value $s=\dh$, meaning that ``fractal rods'' may not even exist for certain fractals. However, due to the simplified continuum structure of fractional theory, ``fractional rods'' matching the power $\a$ of the measure are well-defined, i.e., they yield a positive \emph{finite} result in a measurement. The relation between fractional measures and fractals is not just a parallelism. As proven in a series of theorems \cite{ff1,ff2,ff3,ff4,ff5,ff6,ff7,ff8,ff9}, in $D=1$ a fractional integral of real order $\a$ represents, in a precise quantitative way, the integral over a random fractal with Hausdorff dimension $\dh=\a$; complex-order integrals are in turn associated with integrals over deterministic fractals \cite{NLM}. Fractional spacetimes capitalize on these results and apply their extension to $D\geq 2$ embedding dimensions. Factorizable fractional measures are then interpreted as describing the product $\cF=\cF_0\times\cF_1\times\dots\times \cF_{D-1}$ of $D$ fractals, whose Hausdorff dimension is given exactly by the sum of the dimensions of the sets, $\dh(\cF)=\sum_{\mu=0}^{D-1}\dh(\cF_\mu)$. This equality happens because we are in a smooth geometry, but it does not always hold for product fractals, where in general $\dh(\cF)\geq\sum_\mu\dh(\cF_\mu)$.

Fractional spacetimes can be equipped with generalizations of the Laplace--Beltrami operator \Eq{lbo}. A particular choice is the second-order operator \Eq{ka}, which is self-adjoint with respect to the natural scalar product \cite{frc3}. Laplacian operators with fractional order $2\g\neq 2$ (Ref.\ \citen{frc4}) or second-order but with different measure factors \cite{frc2,frc3} are also possible. 

In the model of multifractional spacetimes with Laplacian \Eq{ka}, momentum space $\cM_w^D=\cM_{\a'}^D$ may be equipped with a different fractional measure $w(p_{\rm frac})=v_{\a'}(p_{\rm frac})$. In the following we dub as $p_{\rm frac}$ the physical momentum in fractional models. The functions
\be 
\bE(p_{\rm frac},x) = \frac{1}{\sqrt{v_{\a'}(p_{\rm frac})v_\a(x)}}\,\frac{\rme^{\rmi p_{\rm frac}\cdot x}}{(2\pi)^{\frac{D}{2}}}
\ee
are eigenfunctions of \Eq{ka}, $\cK_\a\bE(p_{\rm frac},x) = -p_{\rm frac}^2\bE(p_{\rm frac},x)$, where $p_{\rm frac}^2:=\eta_{\mu\nu}p_{\rm frac}^\mu p_{\rm frac}^\nu=-(p_{\rm frac}^0)^2+\sum_{i=1}^{D-1}(p_{\rm frac}^i)^2$. 
 They form a basis for the unitary invertible momentum transform
\bs\ba
\tilde f(p_{\rm frac}) &:=& \int_{-\infty}^{+\infty}\rmd\vr_\a(x)\,f(x)\,\bE^*(p_{\rm frac},x)=:F_{\vr}[f(x)]\,,\label{fo1}\\
f(x) &=& \int_{-\infty}^{+\infty}\rmd\vr_{\a'}(p_{\rm frac})\,\tilde f(p_{\rm frac})\,\bE(p_{\rm frac},x)\,.\label{fo2}
\ea\es
If momentum and position space have the same measure ($\a=\a'$), $F_\vr$ is an automorphism. The identity resolution of this transform is the fractional generalization of the delta distribution, $\de_\a(x,x'):=\de(x-x')/\sqrt{v_\a(x)v_\a(x')}= \int\rmd\vr_{\a'}(p_{\rm frac})\,\bE^*(p_{\rm frac},x)\bE(p_{\rm frac},x')$. Revisiting the discussion on the volume of the $D$-ball, in fractional theory momentum space is defined as the one conjugate to the space spanned by coordinates $x\leftrightarrow p^{-1}_{\rm frac}$, implying that the length of the radius is not $\cR$ but $R=\cR^{1/\a}$. Position measure is defined expressly for coordinates $x$ to possess length units. In other theories such as asymptotic safety and HL gravity, momentum and position space are identified differently, as we shall se below.

Dynamics enters when an action is defined. For example, a scalar field theory is given by $S_\a=\int\rmd\vr_\a(x)[(1/2)\phi\cK_\a\phi-V(\phi)]$ for some potential $V$. The Laplacian also determines the spectral dimension $\ds$ of spacetime, defined as the trace (in position space) per unit volume of the heat kernel operator. It represents the dimensionality felt by a test particle left to diffuse for an infinitesimal time. For natural (i.e., nonfractional) diffusion, one can show that $\ds=\dh$ \cite{frc1}. 

In order to obtain a multiscale geometry, it is sufficient to replaces the measure $\vr_\a$ with a linear superposition of fractional measures, summed over a finite number of parameters $0<\a_n\leq 1$ \cite{fra4,frc2,frc4}:
\be\label{mufm}
\vr(x)=\sum_{n=1}^N g_n\, \vr_{\a_n}(x)\,,
\ee
where $g_n$ are some dimensionful coupling constants containing the scales $\ell_n$ of the system. This type of multicomponent measures is encountered in multifractal geometry and complex systems (see Refs.\ \citen{frc2,frc4} and \citen{Har01} for further details). The diffusion equation is realized by summing over $\a$ and also the order $\b$ of the diffusion operator \cite{fra6,frc4}. The Hausdorff dimension (and also $\ds$, for natural diffusion) is approximately given by $\dh(\ell)\approx D\a_{\rm eff}(\ell)$, where the effective fractional charge reads
\be\label{aeff}
\a_{\rm eff}(\ell)=\frac{1+\sum_{n=1}^{N-1}\zeta_n(\ell)\,\a_n}{1+\sum_{n=1}^{N-1}\zeta_n(\ell)}\,.
\ee
The coefficients $\zeta_n$ have the form $\zeta_1(\ell)=({\ell_1}/{\ell})^2$, $\zeta_n(\ell)=[{\ell_n}/(\ell-\ell_{n-1})]^2$, $\zeta_N\equiv 1$ \cite{fra6,frc4}. Here $\ell=\ell_N$ is the largest scale, which is identified with the probed scale in a measurement. The most important examples are also the simplest ones. When $N=2$, the Hausdorff dimension flows monotonically between two limiting values $\dh\sim D\a_1$ and $\dh\sim D\a_2$. Setting $D=4$, $\a_1=1/2$ and $\a_2=1$, we get a flow from 2 to 4 dimensions. When $N=3$, there are two scales and three values of $\a_n$ which reproduce (for $D=4$, $\a_1=1/2$, $\a_2=1/3$, $\a_3=1$) the asymptotic regimes of the spectral dimension of QEG in the presence of a cosmological constant \cite{fra6,frc4}.

The polynomial measure \Eq{mufm} guarantees that geometry changes with the scale. As a side remark, we notice that it automatically embodies the idea that, in a multifractal spacetime, coordinates acquire a scale dependence \cite{HW,Not08}. Using Eq.\ \Eq{aeff}, one can in fact identify an effective geometric coordinate system (index $\mu$ omitted)
\be\label{qeff}
q_{\rm eff}(\ell) \approx \frac{{\rm sgn}(x)|x|^{\a_{\rm eff}(\ell)}}{\Gamma[\a_{\rm eff}(\ell)+1]} \sim {\rm sgn}(x)|x|^{\frac{\dh(\ell)}{D}}\,,
\ee
which replaces Eq.\ \Eq{geo}. An exact expression for the spectral dimension as well as an exact formula replacing Eq.\ \Eq{qeff} can be found by modifying the \emph{Ansatz} \Eq{mufm} to a factorizable form \cite{frc6}. In particular, the multiscale geometric coordinates simply read
\be\label{3}
q^\mu(\ell)=\sum_{n} g_n(\ell)\, \vr_{\a_n}(x^\mu)\,.
\ee
The substance of the arguments below, which will mainly use Eq.\ \Eq{qeff}, is not changed.

The spectrum of $\cK_\a$ does not differ, as a matter of fact, from that of the standard flat Laplacian, as pointed out in Refs.\ \citen{frc2,frc3,frc4,AIP}, but this is not the only ingredient which can alter the ordinary result $\ds=D$ for the spectral dimension. The explicit calculation can be found, for fixed dimensionality, in Ref.\ \citen{frc1} (where a different second-order Laplacian is used; however, the different distribution of the weight factors to the left and right of the derivatives does not modify the value of $\ds$) and Ref.\ \citen{frc2}. For multifractional spacetimes, the expression of $\ds$ is obtained in Refs.\ \citen{fra6} and \citen{frc4} for an approximate diffusion equation, while an exact multiscale analytic result has been recently presented in Ref.\ \citen{frc6}. The final result for the heat kernel is influenced not only by the spectrum of the Laplacian, but also by (a) the measure structure in position space. Alternatively, if the measure in position space is trivial, it is (b) the structure of momentum space to affect the spectral dimension. The present paper also stresses this point: Namely, that the choice of momentum space strongly affects the physical predictions of a theory on one hand, and, by manipulating it, allows one to map (or, more generically, compare) different models among each other. Case (a) corresponds to the multifractional scenario where the physical momentum is identified with the momentum variable conjugate to coordinates $x$, while case (b) corresponds to models where the position measure is ordinary ($q$ coordinates) but the multiscale dependence is absorbed in the definition of momenta.

Before moving on, we recall some aspects of fractional spacetimes already discussed in previous publications. In (multi)fractional spacetimes, the measure weight $v(x)$ is not dynamical because it describes the differential structure of the underlying geometry/parameter space.\footnote{While in Ref.\ \citen{fra2} $v$ was regarded as a scalar field, in the fractional context the interpretation is different because $v$ is not a Lorentz scalar: it is a given coordinate profile.} The extrinsic geometry, described by the metric, is completely independent. In the presence also of gravity, the measure would become dynamical because it would entail also the determinant of the metric, the latter being one of the dynamical fields of the systems. It is important to stress that the measure $v$ is not supposed to replace the metric determinant; in this paper, the emphasis is placed on a comparison between multifractional spacetimes without curvature and the differential structure (modified Laplacians, and so on) of covariant or partly covariant theories in local inertial frames (Minkowski metric). This is sufficient for all purposes concerning anomalous scaling and effective dimension (which \emph{must} be calculated in inertial frames, even in classical gravity, lest curvature effects vitiate the result).
	
Concerning the structure of the measure itself, factorizability in a product of independent measures is a requirement of utmost importance allowing us to define a momentum transform (and, hence, a consistent momentum space) \cite{frc3} and a consistent quantum mechanics \cite{frc5,frc6} (compare, in contrast, the early nonfactorizable attempt in Refs.\ \citen{fra1,fra2,fra3}). Factorizable measures break all ordinary Poincar\'e symmetries but, on one hand, ordinary symmetries are recovered in the IR at sufficiently large scales (at leasr classically) and, on the other hand, in the UV other symmetries appear which avoid operator proliferation in the theory \cite{frc6}.

As we already mentioned, real-order fractional integrals in the continuum are known to represent approximations of integrals over random fractals, while complex-order integrals represent approximations of integrals over deterministic fractals, thus generalizing the real-order case (random fractals can be taken as ``averages'' of deterministic fractals). Therefore, fractional models can be regarded as an effective/approximate continuum descriptions of field theories living on a generic random or, respectively, deterministic fractal with a certain Hausdorff and spectral dimension. In this sense, no particular form of a single multifractal is chosen and it is clear that every multifractal flat space admits this kind of simple factorized measure. If one wished to include curvature in the picture, in principle it is sufficient to place the fractal on a manifold. Curvature and fractal structures are two independent aspects of geometry.

Once a measure profile $v(x)$ is chosen, however, one can simply forget about the fractal interpretation and regard these continuous spacetimes as anomalous. The theory displays preferred points (such as the singularity of $v$ for a fractional measure). However, the choice of a profile can be regarded as a choice of \emph{presentation}, rather than background. Suppose, for instance, to replace the measure $v(x)\sim |x|^{\alpha-1}$ with another one where the singularity is shifted, $z(x)\sim |x-x_*|^{\alpha-1}$. Then, the results for the Hausdorff and spectral dimensions would be unchanged (even for a rotation-invariant nonfactorizable form $v({\bf x})\sim |{\bf x}|^{(D-1)(\alpha-1)}$ for the spatial part), and the specific aspects of the theory discussed in this paper would not undergo any modification. Things such as the numerical value of the volume of a unit ball do change \cite{frc1} but they are part of the definition, or the conventions, or the presentation of the theory and no physical meaning should be attached to that. On the other hand, in classical and quantum mechanics the ``preferred point'' in the measure modifies the profiles of the dynamical solutions and wavefunctions, respectively \cite{frc5}, but not of probability for free systems. Not only are fractional effects very tiny already at atomic scales and larger \cite{frc2,frc5}, but the relative difference between inequivalent fractional theories is negligible as well \cite{frc1}. Thus, presentation effects seem to play an unobservable role, at least at sufficiently large scales. In general, different choices of measure correspond to inequivalent theories which, however, share the same features starting from anomalous scaling and dimension, the scale dependence of parameter space, and so on \cite{frc1}.

Multifractional spacetimes are models of quantum geometry in two distinct ways. As a fundamental theory, by prescribing a certain action and then quantizing gravity perturbatively on the same footing than other fields. As a phenomenological model, by regarding these spacetimes as effective descriptions of geometry valid in certain regimes of a given fundamental theory. In the second case (which we will implicitly assume in Sec.\ \ref{mfm}) the details of the action are engineered according to the guidelines of the fundamental theory. As anticipated in Sec.\ \ref{intro}, this tailoring procedure constrains the choice of symmetries and of Laplacian and, ultimately, it boils down to a choice of momentum space.


\subsection{Asymptotically-safe quantum gravity}\label{ori}

Quantum Einstein gravity aims to quantize the gravitational interaction as a field theory. Although graviton perturbation theory suffers from divergences, the program can be carried out nonperturbatively via the functional renormalization group approach. This method has wide applications (see Refs.\ \citen{BB,BTW,Pol01,Ros10} for general reviews) but here we only recall the main results in QEG (e.g., Refs.\ \citen{ReS11} and \citen{RSnax} and references therein). 

To uniform our notation with the literature, we use the symbol $k$ for the momentum/energy scale representing a cutoff above which all quantum fluctuations are integrated out. For a given $k$, effective QEG spacetime and dynamics are described by the average action $\Gamma_k[g_{\m\n}]$. The metric is subject to quantum fluctuations and, depending on the resolution $L(k)$ of the microscope probing the geometry, one will observe different properties of spacetime. Thus, the system is multiscale. The change of the microscope amounts to a coarse-graining procedure. Denote as $\langle g_{\mu\nu}\rangle_k$ the metric averaged over Wick-rotated spacetime volumes of linear size $L=L(k)$. Then, $\langle g_{\mu\nu}\rangle_k$ is solution of
\be\label{fe}
\frac{\delta\Gamma_k}{\delta g_{\mu\nu}}\left[\langle g_{\mu\nu}\rangle_k\right]=0\,.
\ee
Letting $k$ take any nonnegative value, one ends up with a continuous family of actions and of field equations \Eq{fe}, all valid simultaneously. From the scale dependence of $\Gamma_k$, one can extract the solution 
$\langle g_{\mu\nu}\rangle_k$ at any scale from the UV to the IR. Although $\langle g_{\mu\nu}\rangle_k$ is typically a smooth classical metric, its change with the scale, regulated by the RG flow, makes effective QEG spacetime a nonsmooth multiscale (in particular, fractal \cite{ReS11}) object with possibly very ``irregular'' geometry.

In the absence of matter and in the so-called Einstein--Hilbert truncation of $\Gamma_k$, the scale dependence of the average metric can be recast in terms of the running cosmological constant $\bar{\lambda}_k$. In fact, the field equations read $R_{\mu\nu}[\langle g\rangle_k]= [2/(2-D)]\bar{\lambda}_k \, \langle g_{\mu\nu}\rangle_k$. Let $k_0$ be an arbitrary reference scale (typically in the infrared, $k_0/k\gg 1$) and assume that the cosmological constant scales according to a function $F$,
\be
\bar{\lambda}_k=F(k^2) \bar{\lambda}_{k_0}\,, 
\ee
where $F$ depends on $k^2$ by the requirement of Lorentz invariance. We also assume that $F$ stays positive throughout the flow between $k$ and $k_0$. Then, from $F^{-1}\,R^\mu_{\;\;\nu}[\langle g\rangle_k]= [2/(2-D)]\bar{\lambda}_{k_0}\,\delta^\mu_\nu$ and the scaling property $R^\mu_{\;\;\nu}[c\,g]=c^{-1}\,
R^\mu_{\;\;\nu}[g]$ of the Riemann tensor ($c>0$ is a constant), one gets
\be\label{metricscaling}
\langle g_{\m\nu} \rangle_k = \frac{1}{F(k^2)} \langle g_{\m\nu} \rangle_{k_0} \, , \qquad 
\langle g^{\m\nu} \rangle_k = F(k^2) \langle g^{\m\nu} \rangle_{k_0} \, .
\ee
In particular, from the inverse metric it follows that the Laplacian $\Delta(k)=(\langle g \rangle_k)^{-1/2}$ $\times\p_\mu(\langle g \rangle_k^{1/2}\langle g^{\m\nu} \rangle_k\p_\nu)$ scales as
\be\label{Laplacescaling}
\Delta(k) = F(k^2) \, \Delta(k_0)\,.
\ee

The function $F$ acquires different asymptotic forms depending on the regime. In the far IR, $F\sim 1$ by definition ($k\to k_0\to+\infty$). In an intermediate semiclassical regime near the Gau\ss{}ian fixed point, $F\sim k^4$. In the deep UV, at the non-Gau\ss{}ian fixed point, asymptotic safety implies that $F\sim k^2$ \cite{LaR5}. All in all, we can write
\be
F\sim |k|^\delta
\ee
in asymptotic regimes (plateaux of the spectral dimension $\ds=2D/(2+\de)$), with $\de=0,4,2$ in the IR, semiclassical limit and UV, respectively. The semiclassical limit may change depending on the truncation and on the presence of matter \cite{RSa}. Outside asymptotic regimes, the details of dimensional flow are nonuniversal and depend on the regularization scheme \cite{frc4,CES}. Notice that the effective spacetime measure is the standard Lebesgue measure $\rmd^Dx\sqrt{|g|}$, so in the zero-curvature limit the Hausdorff dimension coincides with the topological one, $\dh=D$.

In classical geometries (i.e., for nongravitational theories) \cite{BTW}, $\Gamma_k$ can be qualitatively interpreted as encoding  dynamical variables which are coarse-grained on regions of spacetime of size $L$. In this context, in the simplest classical noncompact case $L\sim k^{-1}$ and the resolution is directly identified with $k$. This is the sense in which $\Gamma_k$ acts as a ``microscope'' with resolving power $L(k)$. In quantum gravity not only is the functional form $L(k)$ more complicated \emph{a priori}, but the relation between resolving power, proper distances and IR cutoff $k$ is much subtler. This has been studied in detail in Refs.\ \citen{RSc1} and \citen{RSc2} via a simple algorithm: (i) For every fixed cut-off $k$, determine the eigenfunctions of the Laplacian with eigenvalue $-k^2$. (ii) Along each direction, find the characteristic scale $\Delta x^\mu$ on which these eigenfunctions vary (for instance, for a periodic eigenmode $\Delta x$ is the period); this determines a minimum resolution $\Delta x(k)$ beyond which the geometry becomes ``fuzzy.'' (iii) Define $L(k)=\sqrt{\langle g_{\mu\nu}\rangle_k \Delta x^\mu \Delta x^\nu}$. In a general quantum manifold with fractal properties, the $k$-dependence of lengths is less trivial and encodes the typical effect of momentum dependence of measurements of multifractals. There appear situations where $L$ can decrease arbitrarily even when there exists a non-vanishing minimum resolution: the limit $k\to \infty$ is no longer sufficient to probe arbitrarily small proper distances, because the effective quantum geometry ``shrinks'' faster in the UV limit. This is a direct consequence of the scaling \Eq{metricscaling} of the average metric and, more generally, of the running of the couplings in the effective action. Lengths can be also measured by a macroscopic observer with average metric $\langle g_{\mu\nu}\rangle_{k_0}$, in which case the observer does see a finite minimal length corresponding to the minimal resolution.


\subsection{Ho\v{r}ava--Lifshitz gravity}\label{hlg}

In HL spacetimes, anomalous scaling is associated with the coordinates rather than the metric.
In anisotropic ``critical'' systems, coordinates scale as 
\be\label{as}
t\to \la^z t\,,\qquad {\bf x}\to \la{\bf x}\,,
\ee
for constant $\la$, so that time and space directions have dimensions $[t]=-z$ and $[x^i]=-1$ in momentum units. 
Anisotropic scaling between time and space appears in systems such as Lifshitz scalar field theory \cite{Lif41a,Lif41b,AFF,Car96,CL},
\be\label{f1}
S_{\rm Lifshitz}=\frac12\int \rmd t\rmd^{D-1} x\left[\dot\phi^2-\frac14(\Delta\phi)^2\right]\,,
\ee
where a dot denotes a derivative with respect to time $t$ and $\Delta=\p_i\p^i$ is the spatial Laplacian. In this example, the dynamical critical exponent (or anisotropic scaling exponent) is $z=2$. Its value characterizes the critical behaviour of correlation functions  of the scalar field near a phase transition, as one can infer from the conformal dimension $[\phi]=(D-1-z)/2$. When $D-1=z$ (two spatial dimensions), the field propagator becomes logarithmic and the system is said to be at quantum criticality. Examples of multicritical systems are certain metamagnets, liquid crystals and Ising models. These can be studied with RG techniques; the Lifshitz scalar theory \Eq{f1} possesses two Gau\ss{}ian fixed points, one at $z=2$ where the system is invariant under the anisotropic scaling \Eq{as} and one at $z=1$ where the operator $(\Delta\phi)^2$ becomes irrelevant and local Lorentz invariance can be restored by a relevant operator $\sim \phi\Delta\phi$. 

In the case of gravity \cite{Hor2}, one constructs a $z=D-1=3$ action with all possible relevant operators respecting foliated diffeomorphisms (i.e., spatial rotations and time reparametrizations). We do not write the action because there exist several inequivalent proposals, depending on the assumptions \cite{Hor11}. In general, there are up to $O(\p^{2z})$ operators both in the gravitational and matter sector. The spectral dimension turns out to be $\ds=1+(D-1)/z$ \cite{Hor3,SVW1,CES}. Thus, in the UV $\ds=2$. As the Lifshitz scalar example suggests from the behaviour of the propagator, the advantage in formulating gravity as a critical anisotropic system is that the theory is perturbatively (power-counting) renormalizable. Lorentz invariance is recovered, at least at the classical level (but see Ref.\ \citen{IRS}), in the infrared.


\section{Multifractional Picture of the RG Flow}\label{mfm}

QEG and HL gravity implement anomalous scaling in different ways. On one hand, the formalism of QEG is essentially background-independent and based upon general relativity. Equations are manifestly covariant and the fundamental object is the metric. The latter is argued to carry the anomalous scaling as a direct consequence of the scaling of the cosmological constant in Einstein's equations, Eq.\ \Eq{metricscaling} \cite{RSnax,RSc2}. On the other hand, HL gravity was originally formulated as a generalization to gravity of multicritical systems, where coordinates are assumed to carry an anomalous scaling. Contrary to QEG, this setting bears the imprint of the original background-dependent Lifshitz model, a scalar field theory on a fixed spacetime without curvature, where the metric is a nondynamical object and anomalous scaling is naturally attached to coordinates. The multifractional picture is closer to the HL one, since it is presented in a coordinate-dependent fashion as a field theory. 

Multifractional theory and RG flow are multiscale systems with similar characteristics. The purpose of this section is to play on the fact that this similarity is not accidental and give a complementary description of RG-based theories. The idea is simple and is based on the observation that in the RG flow the physical momentum carries the scale dependence.\footnote{For instance, in perturbative RG a change in the momentum with the scale can be seen as a modification of the kinetic term, corresponding to field-strength (wave function) renormalization.} In multifractional theory, multiscaling is realized in position space by a scale-dependent measure and, hence, effectively scale-dependent coordinates. In turn, this translates into scale-dependent conjugate momenta, which can be compared with the momenta in other RG-based theories.


\subsection{Multifractional RG flow and asymptotic safety}\label{mfmqe}

A first cursory look at multifractional spacetimes often inspires the following crude correspondence with a covariant theory. Comparing multifractional actions with any covariant action, the measure weight $v(x)$ in flat space mimicks the determinant factor $\sqrt{-g}$ on a curved background. Any anomalous scaling of the metric $g_{\mu\nu}$ would survive in a local inertial frame, in which case it can be ascribed, effectively, to a nontrivial measure weight. 

At the effective-action level, for QEG effective spacetimes we can make the argument more precise (on a local inertial frame). 
 We first develop it for \emph{fixed} dimensionality. As we have seen, in $D$ dimensions the QEG average metric $g_{\mu\nu}(k)=\langle g_{\mu\nu}\rangle_k$ scales as
\be\label{gkg} 
g_{\mu\nu}(k) = |k|^{-\delta} g_{\mu\nu}(k_0)\,,
\ee
where $|k|$ is the norm of the vector $k^\mu$ in imaginary time representing the IR cut-off of the theory. Wick-rotating back to Lorentzian signature, the scaling law is presented in a Lorentz-invariant way. Preservation of Lorentz invariance guarantees control over operator proliferation along the RG flow.
Equation \Eq{gkg} implies $\sqrt{-g} = |k|^{-D\delta/2}\sqrt{-g_0}$. From now on, let us identify the RG cutoff scale $k$ with the physical momentum,
\be
k\equiv p_{\normalfont\textsc{qeg}}\,;
\ee
this minimal identification is employed in the literature (but see Ref.\ \citen{CES}) and does not result in a loss of generality for our purposes. In fractional theory, the transform \Eq{fo1} maps functions in position space to functions in physical momentum space, so that the momentum $p_{\rm frac}$ is conjugate to the coordinate $x$, $p_{\rm frac} \leftrightarrow x^{-1}$. If we identified the physical momenta of both theories with each other,
\be\label{ide1}
p_{\normalfont\textsc{qeg}} \stackrel{?}{\sim} p_{\rm frac}
\ee
for each direction, then $k=p_{\normalfont\textsc{qeg}}\leftrightarrow x^{-1}$ and the nontrivial measure in the effective action $\Gamma_k$ would be ``$x^{D\delta/2}$,'' where this writing loosely represents the Euclidean modulus $|x|^{D\de/2}$ or a generic power of the coordinates such as $|x_0\dots x_{D-1}|^{\de/2}$. By itself this point is unclear, but further problems arise on the fractional side, where the measure is $|x_0\dots x_{D-1}|^{\a-1}$. This is a negative power, which could never match with $\de/2\geq 0$.\footnote{If we compared this measure with the effective measure ``$|x|^{-2\delta}$'' of $\Gamma_{k_0}$, we would not fare better. In that case, $\a = 1-\delta/2$, which does not do the job since $\a>0$ but $\delta\geq 2$ in the UV and semiclassical regimes.} To get the correct matching, reproducing the dimensional flow and the associated profile of the spectral dimension, we should set
\be\label{ade}
\a = \frac{2}{2+\delta}\,,
\ee
implying the identification $p_{\normalfont\textsc{qeg}} \leftrightarrow x^{-\a}\sim q^{-1}$ and, therefore,
\be\label{ide2}
{p_{\normalfont\textsc{qeg}} \sim (p_{\rm frac})^\a\,.}
\ee
This relation replaces \Eq{ide1} and, \emph{a posteriori}, is more natural than any other. In fact, $\{q\}$ are nothing but the geometric coordinates \Eq{geo} of fractional spaces. In these special coordinates, the measure becomes an ordinary Lebesgue measure $\rmd^Dq$, with Poincar\'e symmetries acting on $q$ (which are, therefore, nonlinear transformations on $x$). This also fixes the mismatch between the Lorentz-invariant QEG framework and the Lorentz-breaking fractional setting. To carry out the fractional dual picture of asymptotic safety, \emph{we must identify QEG spacetime coordinates with fractional geometric coordinates}.

Equation \Eq{ide2} states that the physical momentum $p_{\normalfont\textsc{qeg}}$ in QEG is mapped to the inverse scale conjugate to $q$. Giving length units to both types of coordinates by associating them with measurement scales,
\be 
x\leftrightarrow p_{\rm frac}^{-1}\sim\ell\,,\qquad q\leftrightarrow p_{\normalfont\textsc{qeg}}^{-1}\sim L\,,
\ee
their role is to define the ``rods'' at a given scale. In fractional theory, $x$ has engineering dimension $-1$ an so $\ell$ is a scale at which measurements are taken with a classical, macroscopic rod. The scaling of $q$ is different, so $L$ is not a length scale but a power of length, $L\sim\ell^\a$. But this is nothing but the ``length'' of a natural rod at the scale where the dimension is (for each direction) $\alpha$. In a theory where physical momentum is $p_{\normalfont\textsc{qeg}}$, it is the $q$'s that have length units.

We can give a pictorial reformulation of this statement, first for a fractal with fixed dimension embedded in an ambient space, and then for a multifractal.
\begin{itemize}
\item {\it Fixed dimension.} Imagine an observer $\cO_\cF$ living in a fractal space $\cF$ and making measurements with a rod. Call the latter a ``$q$-rod,'' measuring lengths in units of ``$q$-meters.'' From the perspective of another observer $\cO$ living outside the fractal, a $q$-meter is the $\alpha$-power of an ordinary ``integer'' meter (or ``$x$-meter,'' or just a ``meter''). \emph{Vice versa}, the $x$-rod of an integer observer $\cO$ measures meters, but from the perspective of the fractal observer $\cO_\cF$ an $x$-rod measures $(1/\a)$-powers of $q$-lengths, not $q$-lengths. If one did not make this correspondence, one would be insisting in measuring the coast of Britain \cite{Man67} with an infinitesimal but ordinary integer rod, getting infinity as a result. 
\item {\it Multiscale geometry.} In the multifractal case, one replaces $\cO_\cF$ with a microscopic observer measuring the fine scales of a multifractal and $\cO$ with a macroscopic observer living at scales of the multifractal large enough so that the set appears smooth ($\dh\sim D$). Suppose the multifractional system is characterized by $N$ charges $\a_n$, corresponding to $N$ asymptotic regimes. For each regime, the geometric coordinates $q_n^\mu$ depend on a given $\a_n$. At scales $\ell\sim \ell_1$, corresponding to QEG scales $L(\ell)\sim L_1=L(\ell_1)$, the QEG momentum is given by Eq.\ \Eq{ide2}, $p_{\normalfont\textsc{qeg}} \sim (p_{\rm frac})^{\a_1}\leftrightarrow q_1^{-1}$; at scales $\ell\sim \ell_2$, $p_{\normalfont\textsc{qeg}} \sim (p_{\rm frac})^{\a_2}\leftrightarrow q_2^{-1}$; and so on. Approximately, all the discussion above and below holds replacing $q$ with the effective scale-dependent coordinate $q_{\rm eff}$, Eq.\ \Eq{qeff}, so that
\be\label{peff}
p_{\normalfont\textsc{qeg}}(L) \sim p_{\rm eff}(\ell) =\frac{{\rm sgn}(p_{\rm frac})|p_{\rm frac}|^{\a_{\rm eff}(\ell)}}{\Gamma[1+\a_{\rm eff}(\ell)]}\  \leftrightarrow\ q_{\rm eff}^{-1}(\ell)\,.
\ee
The duality is the first relation $\sim$ of this equation. In this case, the heuristic relation between $L$ and $\ell$ becomes transcendental, $L\sim \ell^{\a_{\rm eff}(\ell)}/\Gamma[\a_{\rm eff}(\ell)+1]$.
\end{itemize}

In QEG, the RG scaling stems from the comparison of any given scale $1/k=L$ with a classical scale $1/k_0=\ell$. To get finite results, the rod used must be $k$-adapted, yet $k_0$-dependent. Saying that $k=p_{\normalfont\textsc{qeg}}$ is the momentum conjugate to some $q(x)$ is the precise transposition of the idea of getting a $k$-adapted measurement (in fractional language, a $q$-rod, suitable for fractals) but dependent on a classical scale in a certain way motivated by the RG flow (i.e.\ $q$ is a specific function of $x$ given by fractional geometry). Thus, the following correspondences ensue:
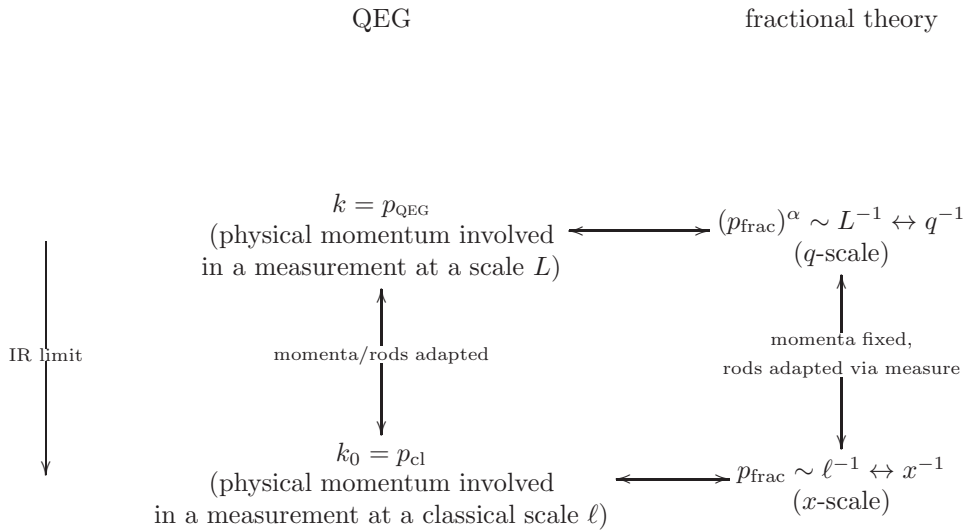
\begin{center}
$$
\xymatrix@R=55pt@C=35pt{& \txt{QEG} & \txt{fractional theory} \\
\ar[d]|*\txt{\scriptsize IR limit} & \txt{$k=p_{\normalfont\textsc{qeg}}$\\ (physical momentum involved\\ in a measurement at a scale $L$)} \ar@{<->}[d]|*\txt{\scriptsize momenta/rods adapted} \ar@{<->}[r] & \txt{$(p_{\rm frac})^\a\sim L^{-1}\leftrightarrow q^{-1}$\\ ($q$-scale)} \ar@{<->}[d]|*\txt{\scriptsize momenta fixed,\\ \scriptsize rods adapted via measure} \\
& \txt{$k_0=p_{\rm cl}$\\ (physical momentum involved\\ in a measurement at a classical scale $\ell$)} \ar@{<->}[r] & \txt{$p_{\rm frac}\sim \ell^{-1}\leftrightarrow x^{-1}$\\ ($x$-scale)}}
$$
\end{center}

Since $L=\ell^\a$ and $\a$ is the Hausdorff dimension (along one direction), one concludes that \emph{the very concept of dimensional flow is, in fact, the notion of adapted rod}. This was already appreciated in the past at least in the case of QEG \cite{RSc1,RSc2}. As we recalled at the end of Subsec.\ \ref{ori}, measuring probes are naturally adapted with the scale in asymptotic safety, where one can construct precise relations among a proper length $L(k)$, the IR cutoff $k$, and the fuzziness of quantum spacetimes. The QEG coordinate system is $k$-independent, and so are also diffeomorphisms; a continuous scale hierarchy is established from the average metric, and an IR reference scale $k_0$ is assumed. All this stems solely from the functional RG approach applied to gravity, and is actually independent from the assumption of asymptotic safety. From the perspective of fractional spaces, the hierarchy is realized via a scale-dependent measure, which in turn defines composite scale-dependent coordinates $q(x,\ell)$. The above diagram illustrates how the two cases can be mapped one onto the other.

Notice that, in fractional theory, the nontrivial measure $\vr_\a(x)$ is not a background but the prescription of the rods by which one makes measurements during the RG flow. At least in the asymptotic regimes of constant spectral dimension, renormalization group and fractal geometry are just two different languages describing a similar geometry (not the same, since symmetries may be different; see below). In QEG one does not change the measure but changes the momenta (hence the rods), while in multifractional spaces one does not change the momenta but changes the measure (hence the rods again): the two pictures are complementary. Recalling the above analogy of the fractal examined by ``fractal'' and ``integer'' observers, one case (fractional theory) corresponds to the integer observer $\cO$ measuring the fractal (at any of its scales) with his own $x$-rods, while the other (QEG) corresponds to the fractal observer $\cO_\cF$ measuring $\cF$, at all scales (see Table \ref{tab1}). 
\begin{table}[ht]
\tbl{QEG versus multifractional picture.}
{\begin{tabular}{|l||c;{2pt/2pt}c|c|}\hline
                     & \multicolumn{2}{c|}{Multifractional spacetimes} & QEG spacetimes \\ \hline
                     &                & Mapping               &                          \\\hline\hline
Coordinate           & $x$            & $q(x,\ell)$ & $q$                      \\\hdashline
Physical momentum    & $p_{\rm frac}$ & $p(\ell)$   & $p_{\normalfont\textsc{qeg}}$         \\\hline
Scale dependence     & implicit       & explicit   						& implicit                 \\\hline
Probed scale         & $\ell= p_{\rm frac}^{-1}$ & $L= p^{-1}(\ell)$ & $L= p_{\normalfont\textsc{qeg}}^{-1}$ \\\hline
Rods adapted via     & measure        & momenta & momenta \\\hline
Laplacian            & $\cK_\a\sim \p^2_x$ & & $\B\sim \p^2_q$\\\hline
\end{tabular}
\label{tab1}}
\end{table}

By definition, physical momentum changes according to the scale of the experiment ($p_{\normalfont\textsc{qeg}}\sim L^{-1}$, $p_{\rm frac}\sim \ell^{-1}$), but a scale dependence hidden in the momentum/coordinate structure is always \emph{implicit}. The mapping to multifractional theory makes this scale dependence explicit. To summarize, (a) in QEG the use of $q$-rods implies that the relation between momenta $p_{\normalfont\textsc{qeg}}$ and probed scales $L$ is fixed throughout dimensional flow and, as a consequence, the Hausdorff dimension of QEG spacetime does not change, $\dh=D$; (b) in multifractional spacetimes, the relation between momenta $p_{\rm frac}$ and probed scales $\ell$ is also fixed, but the measure changes, which leads to a nonconstant Hausdorff dimension $\dh\leq D$; (c) Writing multifractional theory with coordinates \Eq{qeff} and momenta \Eq{peff} (or the exact expression \Eq{3} and its conjugate momentum) reproduces the QEG situation because the measure is trivialized and scale dependence is absorbed and made \emph{explicit} into the coordinate system.

Lorentz-symmetry violation is manifest in multifractional theory\footnote{If one attempted to work with a measure preserving as many Lorentz symmetries as possible, for instance rotations ($v_\a=|{\bf x}|^{\a-1}$), one would obtain a model with the correct anomalous scaling but which is technically untractable after a few steps \cite{fra1,fra2,fra3}. There is no obvious (and perhaps even nonobvious) momentum transform for nonfactorizable measures, the propagator of fields is difficult to compute, and so on. The factorized form \Eq{fac} is crucial.} but not usually expected in asymptotic safety. The mapping, however, was done at the level of the action measure between geometric coordinates $q$ in multifractional theory and momenta $p_{\normalfont\textsc{qeg}}$ in QEG. The measure in geometric coordinates is Lorentz invariant, $\rmd^D q$, but the multifractional Laplace--Beltrami operator is not Lorentz invariant either in $x$ or in $q$ coordinates, due to the weight factors, and differs from its QEG counterpart (see the last line of Table \ref{tab1}, where we made use of Eqs.\ \Eq{effas} and \Eq{ade}). Hence the two theories are physically different although their dimensional flows share a common description.


\subsection{Multifractional RG flow and Ho\v{r}ava--Lifshitz gravity}

To mimick the results of HL gravity, we look at an anisotropic fractional model with
\be\label{hlmap}
\a_0=1\,,\qquad \a_i=\frac1z=\frac1{D-1}\,,\qquad i=1,\dots,D-1\,.
\ee
The spatial fractional charge $\a_i$ is nothing but the inverse of the critical exponent $z$. In particular, in four dimensions $\a_i=1/3$. The coordinates correspondence runs as
\be
x^0_{\rm frac}=x^0_{\rm HL}=t\,, \qquad q^i=x^i_{\rm HL}\,.
\ee
The measure then reads $\rmd\vr_\a=\rmd\vr_{\rm HL}=\rmd t\,\rmd^{D-1} {\bf q}$. Physical momenta are defined consequently: $p^0_{\rm frac}=p^0_{\rm HL}$, $(p_{\rm frac}^i)^\a\sim p^i_{\rm HL}$. 

To get a multiscale geometry, one applies perturbative RG considerations to build the total action with a hierarchy of differential Laplacian operators, from order $2z$ (UV) to $2$ (IR). While in HL gravity the UV spectral dimension is anomalous due to the higher-order Laplacian, in the fractional mapping it is so because of the nontrivial fractional charges.

Again, symmetries and action differ with respect to multifractional theory taken as a stand-alone proposal, so the physics is ultimately inequivalent. In the latter case, the Laplace-Beltrami operator is isotropic, Eq.\ \Eq{ka}, $\cK_\a\sim-p^2_t+\p^2_x$, while the operator corresponding to HL spacetime in the UV is $\cK\sim-\p_t^2+\p_q^{2/\a}$. This is the reason why, in particular, in $D=2+1$ dimensions the two models predict a different UV spectral dimension, $\ds\sim 3/2$ (as in QEG) in the fractional case and $\ds=2$ in the HL case. Our purpose here was to show how the main geometric features of HL dimensional flow can be easily reproduced by the multifractional framework.


\section{Discussion}\label{disc}

A twofold purpose of multifractional geometry is, when regarded as an independent theory, to provide a general analytic explanation of dimensional flow \cite{frc4} and, when regarded as an effective model, to reproduce the flow in other theories and give a complementary description. Both aims can be stated and reached in such a generality because multifractional spacetimes are constructed with the wide-scope tools of multifractal geometry, anomalous transport and complex systems. We have applied this \emph{passe-partout} spirit to the multiscale behaviour of QEG and Ho\v{r}ava--Lifshitz effective spacetimes. In particular, we have revisited their RG flow within fractal geometry.

It is worth stressing again a point made in Subsec.\ \ref{mfmqe}. The mapping/comparison between fractional theories and Lorentz-invariant or partially Lorentz-invariant theories is done at the level of parameter space (coordinates and momenta), not at the level of the Laplacian/action symmetries. Then, the breaking of Poincar\'e symmetries in multifractional theories lies out of the main focus of the paper. In this context, multifractional models are used only in relation to their measure structure and not to the Laplacian structure, at which point one recognizes the intrinsic difference among the models. Consequently, one can start from, say, asymptotic safety, and recast its phase-space structure (i.e.\ position and momentum coordinates) in terms of scale-dependent objects, which are then broken down to phase-space coordinates recognized to describe a multifractional spacetime. In practice, the relation is made quantitative by identifying the position-space coordinates ``$x_{\normalfont\textsc{qeg}}$'' of asymptotic safety as the geometric coordinates $q$ of multifractional spaces, for which the measure is the usual one $d^Dq$. By writing $x_{\normalfont\textsc{qeg}}=q(x)$ as a function of the nongeometric multifractional coordinates $x$, as in Eqs.\ \Eq{qeff} and \Eq{3}, one is simply breaking down the parameter space of asymptotic safety into a presentation-dependent language which is convenient to describe anomalous scaling and multifractal-like geometries. Then, one can compare the multiscale flow of multifractional spacetimes with the Lorentz invariant asymptotic-safety case (HL gravity is a somewhat intermediate case where mixed coordinates are used), and we are able to describe quantum geometries of other putative fundamental theories in the language and with the tools of multifractal geometry.

In general, the multiscale nature of the renormalization group is akin to the structure found in multifractal systems.\footnote{Whether there is a deeper link between fractals and anomalous diffusion on one hand and the hidden tree structure of Feynman graphs (Refs.\ \citen{Zim69,Kre00,Gal07} and references therein) on the other hand is a matter of speculation.} At the level of the Laplacian operator, the correspondence between different theories is based on the following heuristic observation. In the UV, the spectral dimension of multifractional, QEG and HL spacetimes is 2 in four dimensions. Calling $\g_\mu$ half the order of the Laplacian along the direction $\mu$ and allowing for a nontrivial position-space measure, the spectral dimension for a nonfractional diffusion operator is $\ds=\sum_\mu \a_\mu/\g_\mu$ \cite{frc4}. This formula is invariant under the interchange
\be
\a_\mu\ \longleftrightarrow\ \frac{1}{\g_\mu}\,,
\ee
mapping theories with $(\a_\mu,\g_\mu)=(1,\g_\mu)$ (ordinary measure, higher-order Laplacians) to a fractional spacetime with $(\a_\mu,\g_\mu)=(1/\g_\mu,1)$ (nontrivial measure, second-order Laplacian). This is realized by Eq.\ \Eq{ade} in asymptotic safety and by Eq.\ \Eq{hlmap} in HL gravity. Despite this intriguing degeneracy, when one looks at the details of each Laplacian one finds differences in the symmetries. Although these are intrinsic and cannot be bypassed, it is nonetheless desirable to clarify the correspondence in the dimensional flow and recast it without invoking the form of the Laplacian. Here we did so and pointed out how a more precise quantitative comparison between quantum field theory renormalization and fractal geometry is based upon the identification of coordinates, measure in position space, and physical momenta. 

As shown in Subsec.\ \ref{mfmqe}, the natural coordinates in QEG do not correspond to the $x$ variables in fractional spacetime but to the ``geometric'' coordinates $q$. This implies that the physical momentum of QEG is the same as in noncommutative geometries with fractal properties, although noncommutative space is spanned by $\{x\}$, not by $\{q\}$, and with a noncommutative product rule \cite{ACOS}. Thus, the physical picture of these noncommutative geometries is somewhat hybrid between quantum gravity and fractional spacetimes. Interestingly, this observation pinpoints one of the key differences between all these independent approaches to quantum geometry: namely, \emph{the choice of momentum space}. In other words, what distinguishes one model from another is, apart from the symmetries, the scale identification. This is also the leverage point where one can start drawing a quantitative mapping between these theories, as shown here and in Ref.\ \citen{ACOS}. A better understanding of these relations may be a promising line of research in the near future.

\section*{Acknowledgments}
The author thanks A.\ Eichhorn, V.\ Rivasseau, F.\ Saueressig and J.\ Th\"urigen for useful discussions. This work is under a Ram\'on y Cajal tenure-track contract.



\end{document}